\documentclass[%
aip,
jmp,
amsmath,amssymb,
preprint,
]{revtex4-1}

\usepackage{graphicx}
\usepackage{dcolumn}
\usepackage{bm}

\newcommand{\ltsim}{\protect\raisebox{-0.5ex}{$\:\stackrel{\textstyle
<}{\sim}\:$}} 
\newcommand{\gtsim}{\protect\raisebox{-0.5ex}{$\:\stackrel{\textstyle
>}{\sim}\:$}} 

\begin{document}

\preprint{AIP/123-QED}

\title[Magnetic Reconnection under Anisotropic MHD Approximation]
{Magnetic Reconnection under Anisotropic MHD Approximation}

\author{K. Hirabayashi}
\email{hirabayashi-k@eps.s.u-tokyo.ac.jp}
\author{M. Hoshino}

\affiliation{ 
Department of Earth and Planetary Science, The University of Tokyo,
7-3-1 Hongo, Bunkyo-ku, Tokyo, Japan
}

\date{\today}

\begin{abstract}
 We study the formation of slow-mode shocks in collisionless magnetic
 reconnection by using one- and two-dimensional collisionless MHD codes
 based on the double adiabatic approximation and the Landau closure
 model.
 We bridge the gap between the Petschek-type MHD reconnection model
 accompanied by a pair of slow shocks and the observational evidence of
 the rare occasion of {\it in-situ} slow shock observations.
 Our results showed that once magnetic reconnection takes place, a
 firehose-sense ($p_{||}>p_{\perp}$) pressure anisotropy arises in the
 downstream region, and the generated slow shocks are quite weak
 comparing with those in an isotropic MHD. 
 In spite of the weakness of the shocks, however, the resultant
 reconnection rate is $10-30\%$ higher than that in an isotropic case.
 This result implies that the slow shock does not necessarily play an
 important role in the energy conversion in the reconnection system,
 and is consistent with the satellite observation in the Earth's
 magnetosphere.
\end{abstract}



\maketitle

 \section{\label{sec:intro}Introduction}
 
 Magnetic reconnection has been widely studied as an efficient mechanism
 for converting magnetically stored energy to thermal and/or kinetic
 energy in plasmas.
 It appears in various phenomena in astrophysics and space physics, e.g.,
 the magnetopause where the Earth's magnetic field reconnects with the
 interplanetary magnetic field carried by the solar wind,
 \cite{Sckopke1981,Hesse1994,Frey2003} the magnetotail where the field
 lines stretched by the solar wind reconnect with each other,
 \cite{Oieroset2001,Nagai2001,Zenitani2012} and solar flares where
 enormous energy and mass are released by magnetic
 reconnection.\cite{Parker1963,Kopp1976,Masuda1994}
 The common feature of these plasmas is that the magnetic Reynolds
 numbers are extremely large and classical Ohmic dissipation cannot
 explain the rates of energy release.\cite{Sweet1958,Parker1957}
 To overcome this problem, Petschek pointed out that the reconnection
 accompanied by a pair of slow shocks emanating from the X-type
 diffusion region could provide a fast reconnection rate even in plasmas
 with high Reynolds numbers.\cite{Petschek1964} 

 In fact, numerous researchers have demonstrated the fast reconnection
 with a pair of slow shocks in MHD nonlinear simulations.
 However, it is rare for the formation of the Petschek-type slow shock
 to be detected in current sheets by {\it in-situ}
 observations.\cite{Saito1995,Seon1996} 
 Furthermore, the recent advances in computational power have enabled
 particle-in-cell and hybrid simulations of collisionless plasmas,
 \cite{Liu2011a, Higashimori2012, Weng2012, Hoshino1998} but the results
 do not provide clear evidence for the formation of the slow shock.
 It is thought that the absence of shocks in kinetic simulations might
 be caused by the smallness of the simulation box or by the pressure
 anisotropy due to the PSBL (plasma sheet boundary layer) ion beams
 accelerated along the magnetic field lines from the diffusion region.
 \cite{Higashimori2012,Liu2011a}
 In fact, a large pressure anisotropy with $p_{||}>p_{\perp}$ has been
 observed during reconnection in the magnetotail\cite{Hoshino2000}.

 In this paper, in order to bridge the gap between the Petschek model
 with its slow-mode shocks and the kinetic simulation results and
 observations, we perform a series of collisionless MHD simulations,
 paying special attention to the effect of a pressure anisotropy on
 the global dynamics of a reconnection layer. 
 We use the MHD equations combined with an anisotropic pressure tensor,
 the double adiabatic approximation, and the Landau closure model as a
 basis for collisionless fluid formulation.

 The outline of this paper is as follows.
 In Section \ref{sec:model}, we briefly introduce the basic equations
 for collisionless MHD and explain the setups of our simulations.
 Section \ref{sec:result} shows the results of one- and two-dimensional
 simulations, i.e., the 1-D result in Section \ref{subsec:riemann} and
 the 2-D result in Section \ref{subsec:2d}.
 The concluding Section \ref{sec:discussion} includes the summary and
 implications of the results of this paper.

 \section{\label{sec:model}Basic Equations and Simulation Models}

  \subsection{\label{subsec:eqs}Collisionless MHD Equations}

  In the MHD limit in which all scales of fluctuations are much larger
  than the ion Larmor radius and all frequencies are much lower than the
  ion cyclotron frequency, the governing equations of a collisionless
  plasma can be described by the Kulsrud's formulation.\cite{Kulsrud1983}
  The basic equations we adopted are as follows:
  \begin{equation}
   \frac{\partial \rho}{\partial t}
    + {\bm \nabla} \cdot \left( \rho {\bm V}\right) = 0,
    \label{eq:mass}
  \end{equation}
  \begin{equation}
   \frac{\partial}{\partial t}\left(\rho{\bm V}\right)
    + {\bm \nabla} \cdot
    \left[ \rho{\bm V}{\bm V}
     +\left({\bm P}+\frac{B^2}{8\pi}{\bm I}\right)
     - \frac{{\bm B}{\bm B}}{4\pi}\right] = 0,
    \label{eq:moment}
  \end{equation}
  \begin{equation}
   \frac{\partial {\bm B}}{\partial t}
    = {\bm \nabla} \times 
    \left( {\bm V} \times {\bm B}\right)
    - {\bm \nabla} \times \left( \eta {\bm \nabla}\times{\bm B}\right),
    \label{eq:induction}
  \end{equation}
  \begin{equation}
   {\bm P} = p_\perp {\bm I}
    + \left( p_{||} - p_\perp \right) {\hat{\bm b}}{\hat {\bm b}},
    \label{eq:p_tensor}
  \end{equation}
  where $\rho$ is the mass density, ${\bm V}$ is the balk velocity,
  ${\bm B}$ is the magnetic field vector,
  ${\bm I}$ is the unit tensor, $\eta$ is the magnetic diffusivity,
  $\hat{\bm b} = {\bm B}/B$ is the unit vector parallel to the magnetic
  field, and ${\bm P}$ is the pressure tensor with the different
  components $p_{\perp}$ and $p_{||}$, which are respectively
  perpendicular and parallel to the background magnetic field.
  
  Equations of state for determination of $p_{\perp}$ and $p_{||}$ are
  given below.
  Taking the second moments of a drift kinetic equation leads to a set
  of equations of state as follows\cite{Chew1956}:
  \begin{equation}
   \rho B \frac{D}{Dt} \left( \frac{p_{\perp}}{\rho B} \right)
    = - {\bm \nabla} \cdot {\bm q}_{\perp}
    - q_{\perp}{\bm \nabla} \cdot {\hat {\bm b}},
    \label{eq:perp0}
  \end{equation}
  \begin{equation}
   \frac{\rho^3}{B^2} \frac{D}{Dt}
    \left( \frac{p_{||} B^2}{\rho^3} \right)
    = - {\bm \nabla} \cdot {\bm q}_{||}
    + 2q_{\perp} {\bm \nabla} \cdot {\hat {\bm b}},
    \label{eq:para0}
  \end{equation}
  where ${\bm q}_{\perp,||}=q_{\perp,||}{\hat {\bm b}}$ are the heat
  fluxes along the magnetic field and
  $D/Dt=\partial/\partial t + {\bm V}\cdot{\bm \nabla}$ is the fluid
  Lagrangian derivative. 
  It can be seen from eqs.~(\ref{eq:perp0}) and (\ref{eq:para0}) that
  if the effects of heat fluxes are neglected (the so-called double
  adiabatic limit or the CGL approximation), the quantities enclosed in
  parentheses on the left-hand sides are conserved and then
  $p_{\perp} \propto \rho B$ and $p_{||} \propto \rho^3B^{-2}$.
  When considering magnetic reconnection, it appears, therefore, that
  the reduction of magnetic field strength and the increase of plasma
  density across the plasma sheet boundary naturally leads to pressure
  anisotropy with $p_{||}>p_{\perp}$.  
  
  For a model of the heat flux, we employed the Landau closure model,
  \cite{Hammett1990,Hammett1992,Snyder1997} which can 
  correctly describe the kinetic linear Landau damping process.
  The functional forms of the linearized heat flux in Fourier space can
  be written as 
  \begin{equation}
   {\tilde q}_{||} = - \sqrt{\frac{8}{\pi}} \rho_0 c_{||0}
    \frac{ik_{||} \left(p_{||} / \rho \right)}{\left|k_{||}\right|},
    \label{eq:q_para0}
  \end{equation}
  \begin{eqnarray}
   {\tilde q}_{\perp}
    &=& -\sqrt{\frac{2}{\pi}} \rho_0 c_{||0}
    \frac{ik_{||} \left( p_{\perp} / \rho \right)}{\left|k_{||}\right|}
    \nonumber \\
   &&+ \sqrt{\frac{2}{\pi}}c_{||0}\frac{p_{\perp 0}}{B_0}
    \left( 1-\frac{p_{\perp 0}}{p_{|| 0}} \right)
    \frac{ik_{||}B}{\left|k_{||}\right|},
    \label{eq:q_perp0}
  \end{eqnarray}
  where subscript $0$ denotes the equilibrium values,
  $c_{||0}=\left(p_{||0}/\rho_0\right)^{1/2}$ is the parallel thermal
  velocity, and $k_{||}$ is the parallel wavenumber.
  The first terms in both eqs.~(\ref{eq:q_para0}) and (\ref{eq:q_perp0})
  represent the effect of heat conduction, and the second term in
  eq.~(\ref{eq:q_perp0}) shows the flux due to the magnetic field
  gradient, which is important for correctly reproducing the threshold
  of mirror instabilities. 

  Real space expressions of these closure equations are given by
  convolution integrals along the magnetic field line. 
  In this paper, to avoid the expensive routine of performing Fourier
  transformation along tangled field lines, we follow the same procedure
  by Sharma et al.\cite{Sharma2006}
  We pick out one characteristic wavenumber $k_L$ along the field line
  which represents the length scale of collisionless linear Landau
  damping and adopt the local approximation in Fourier space.
  Hence $ik_{||}$ and $|k_{||}|$ in eqs.~(\ref{eq:q_para0}) and
  (\ref{eq:q_perp0}) are replaced by 
  $\nabla_{||} = {\hat {\bm b}}\cdot{\bm \nabla}$ and $k_L$,
  respectively. 
  
  Since a finite electric conductivity is assumed for this reconnection
  study, Ohmic heating terms should be added explicitly in the
  right-hand sides of eqs.~(\ref{eq:perp0}) and (\ref{eq:para0}). 
  In the anisotropic MHD, however, the ratio of energy distribution to
  parallel and perpendicular thermal energy cannot be self-consistently
  determined.
  Here we assume that Ohmic heating is mainly caused by electrons rather
  than ions, and that the colliding electrons with waves are rapidly
  isotropized.
  Hence the same amount of energy is deposited into each degree of
  freedom by
  \begin{equation}
   \frac{d}{dt}
    \left( \frac{p_{||}}{\gamma_{||} -1} \right)_{\rm Ohm} \!\!\!
    = \frac{f_{||}}{f_{||}+f_\perp}\eta J^2,
    \label{eq:ohm1}
  \end{equation}
  \begin{equation}
    \frac{d}{dt}
    \left( \frac{p_{\perp}}{\gamma_{\perp} -1} \right)_{\rm Ohm} \!\!\!
    = \frac{f_\perp}{f_{||}+f_\perp}\eta J^2,
    \label{eq:ohm2}
  \end{equation}
  where $f_{||}=1$ and $f_{\perp}=2$ are the degree of freedom  parallel
  and perpendicular to the local magnetic field lines, respectively.
  $\gamma_{||,\perp}=(f_{||,\perp}+2)/f_{||,\perp}$ are adiabatic
  indices.

  \subsection{\label{subsec:setup}Simulation Setup}

  In our simulation study, we solve the continuity equation
  (\ref{eq:mass}), the momentum conservation equation (\ref{eq:moment}),
  the induction equation (\ref{eq:induction}), the equations of
  state (\ref{eq:perp0}) and (\ref{eq:para0}) combined with Ohmic
  heating terms (\ref{eq:ohm1}) and (\ref{eq:ohm2}), and the heat flux
  (\ref{eq:q_para0}) and (\ref{eq:q_perp0}) under the Fourier-space
  local approximation. 

  As an initial state, we assume a static, isothermal, and isotropic
  plasma with the Harris-type current sheet:
  \begin{equation}
   B_X = B_{X0} \tanh(Z/L),
  \end{equation}
  \begin{equation}
   p_{||} = p_{\perp} = p_0
    + \frac{B_{X0}^2}{8\pi} \cosh^{-2}(Z/L).
  \end{equation}
  The subscript 0 denotes the lobe region, and the plasma beta in the
  lobe region is set at $0.25$.
  Since the magnetic field strength appears in the denominator when
  determining the direction of the magnetic field and calculating the
  heat flux, we discuss the guide-field reconnection with a finite
  uniform guide field component $B_Y$ so as to avoid zero division near
  the neutral sheet. 
  The strength of the guide field is characterized by $\phi$, which is the
  angle between the tangential magnetic field and the X-axis, i.e., 
  $\tan\phi=B_Y/B_{X0}$. $\phi=0^\circ$ corresponds to a purely
  anti-parallel case, and the $\phi=90^\circ$ case has no anti-parallel
  component. 
  
  The spatial scale is normalized by the half thickness of the current
  sheet, $L$, and the velocity is normalized by the ${\rm
  Alfv\acute{e}n}$ velocity defined in the lobe region far from the 
  X-point, $V_A$.
  The time scale is normalized by the ${\rm Alfv\acute{e}n}$ transit
  time, $L/V_A$.
  The magnetic pressure and gas pressure are normalized by twice the
  lobe magnetic pressure, $\left(B_{X0}^2+B_{Y0}^2\right)/4\pi$.
  The background magnetic Reynolds number is set to $Re=500$ and
  the anomalous resistivity with $Re=80$ is assumed at the origin.
  We solved the region $X=0, \ -200<Z<200$ with $4000$ grid points in 1-D
  simulations and $0<X<500, \ -50<Z<50$ with $5000 \times 1000$ grid
  points in 2-D simulations.
  In 1-D simulations, we set the normal component of the magnetic field
  $B_Z$ to $5\%$ of the total magnetic field in the lobe region.
  We imposed the symmetric boundary condition on $X=0$ and the free
  boundary conditions on other boundaries. 
  
  The basic equations are all discretized in space and time.
  Spatial derivatives are calculated by the 4th-order central differences,
  and temporal derivatives are integrated by the 4th-order Runge-Kutta
  scheme.
  A numerical error of ${\bm \nabla}\cdot{\bm B}$ is removed with the
  method used in Rempel et al.\cite{Rempel2009}
  Furthermore, for the purpose of capturing shock waves, an artificial
  diffusivity is imposed upon all MHD variables at every time step.
  \cite{Rempel2009}
  
  \section{\label{sec:result}Collisionless MHD Simulation Results}
 
  \subsection{\label{subsec:riemann}1-D Riemann Problem}

  If the steady-state reconnection can be realized in two-dimensional
  space, we may discuss the structure of reconnection as a solution of
  the one-dimensional Riemann problem\cite{Tsai2006,Heyn1988}.
  Before proceeding to the 2-D results, we shall look at the qualitative
  characteristics of reconnection layers in the anisotropic MHD.
  The heat flux is set to zero in this subsection in order to
  clearly observe a discontinuity.
  We note that the presence of the heat flux does not change the
  fundamental reconnection structure.

\begin{center}
   \begin{figure*}
    \includegraphics{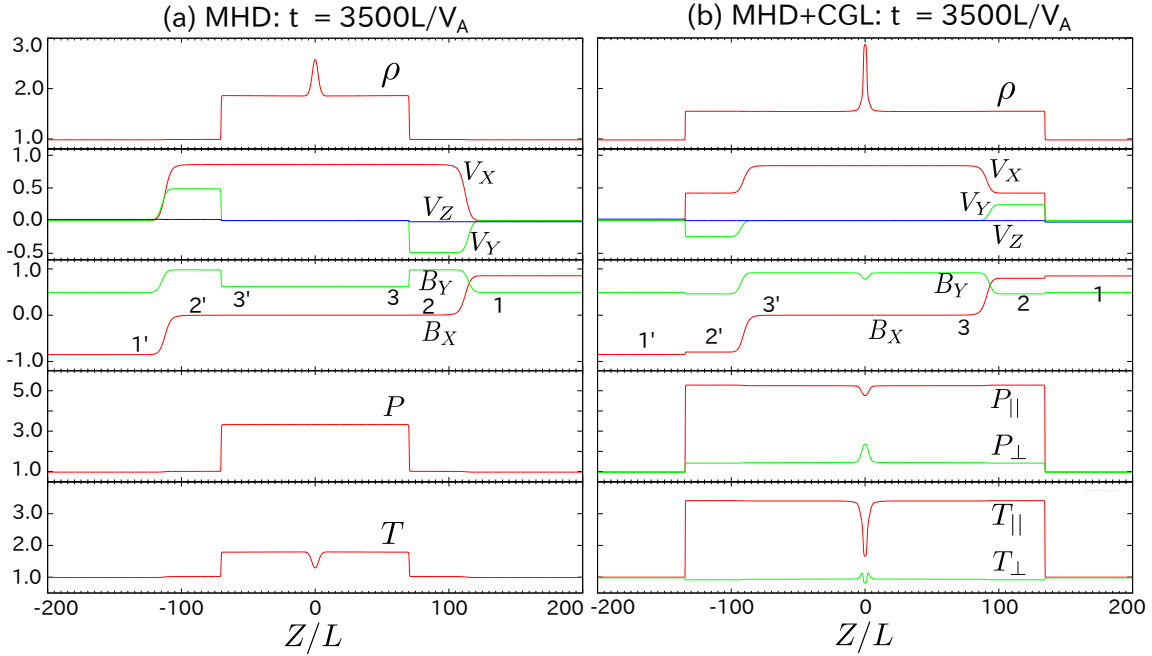}
    \caption{\label{fig:figure1}The spatial distribution of variables at
    time $t=3500L/V_A$ with $\phi=30^{\circ}$. (a)MHD, and (b)MHD+CGL.}
   \end{figure*}
  \end{center}

  Shown in Fig. \ref{fig:figure1} are the results with $\phi=30^\circ$,
  the isotropic MHD case Fig. \ref{fig:figure1}(a) and the MHD+CGL case
  Fig. \ref{fig:figure1}(b).
  The most striking difference between the isotropic and anisotropic
  cases is the sequence of propagation of slow-mode waves and
  ${\rm Alfv\acute{e}n}$ waves or rotational waves, which can be seen
  clearly by comparing the velocity profiles.
  As we have expected, the parallel pressure is more highly enhanced
  across the slow shock than the perpendicular pressure in the CGL case.
  From the next subsection, we discuss the isotropic and anisotropic
  results one by one in order to clearly understand the differences.

  First, we shall discuss the results in the isotropic MHD case.
  In all calculations, first of all, a pair of the fast rarefaction
  waves propagate away from the simulation domain by the time about
  $t=200-300L/V_A$.
  If the lobe magnetic field lines are completely anti-parallel, a pair
  of switch-off slow shocks should propagate outward from the neutral
  point.
  In the case of the guide field reconnection, however, we can clearly
  see the propagation of the rotational discontinuities ahead of the
  slow shocks, which are regarded as finite amplitude ${\rm
  Alfv\acute{e}n}$ waves.   
  Since the initial magnetic field configuration violates the
  coplanarity, and since the slow shocks alone cannot connect both lobe
  regions consistently, the rotational waves are generated so as to maintain
  the coplanarity downstream.
  
  Behind the rotational discontinuity, we can find the slow shock,
  where the density, the pressure, and the temperature increase.
  We find that the compression ratio of density is 1.88 across the shock.

\begin{center}
   \begin{figure*}
    \includegraphics{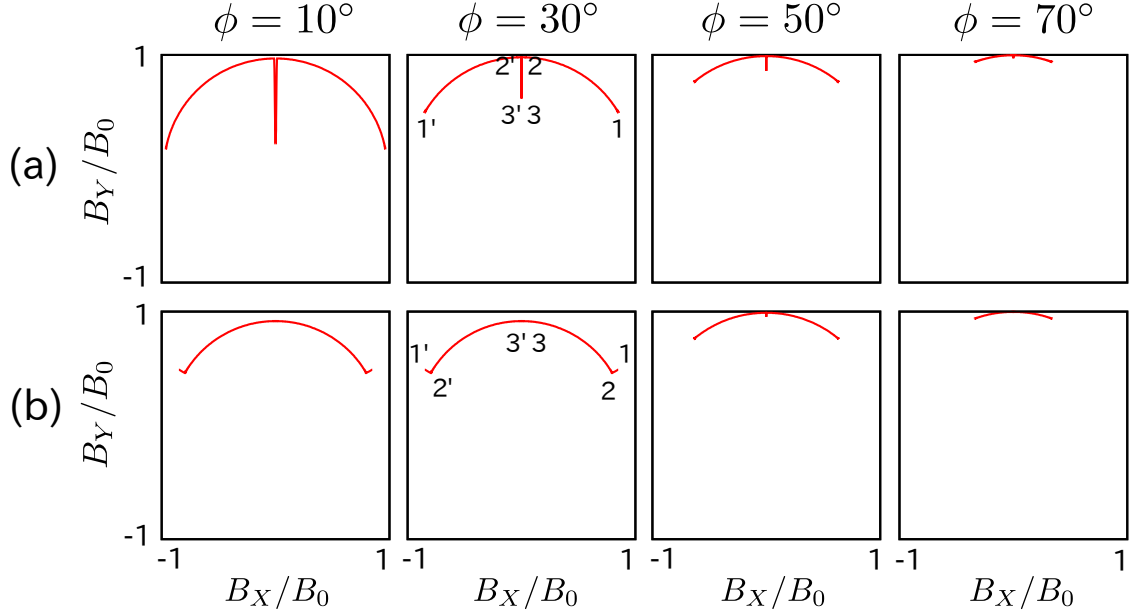}
    \caption{\label{fig:figure2}The hodograms of magnetic fields at time
    $t=3500L/V_A$. (a)MHD case, and (b)MHD+CGL case.}
   \end{figure*}
  \end{center}

  A hodogram is useful to understand the behavior of the
  discontinuities.
  Fig. \ref{fig:figure2}(a) shows the magnetic field in the isotropic
  MHD case with $\phi=10^\circ, 30^\circ, 50^\circ,$ and $70^\circ$.
  In order to focus on the shock structure, the data in the
  region of $|Z|<10L$, where the effect of the initial current sheet
  remains, are removed in the hodograms.
  Numbers in the panel for $\phi=30^\circ$ indicate the spatial
  correspondence with Fig. \ref{fig:figure1}.
  When crossing the rotational discontinuity, the trajectory draws a
  circle with the center at the origin (1-2,1'-2'), which means the
  magnetic field changes its direction with the magnetic field energy
  conserved.
  When a slow shock passes, on the other hand, the distance from the
  origin in the hodogram decreases (2-3,2'-3'), which corresponds to the
  reduction of the magnetic energy.
  In Fig. \ref{fig:figure2}(a), these reductions appear as vertical
  trajectories along the $B_Y$ axis.
  As the shear angle $\phi$ increases, the vertical displacement becomes
  small (i.e., decrease in released magnetic energy), but the topological
  structure of a reconnection layer does not change.\cite{Tsai2006}

  Next, we shall look at the anisotropic case of
  Fig. \ref{fig:figure1}(b).
  After a pair of fast rarefaction waves propagate away by the time
  about $t=200-250L/V_A$,
  the slow shocks appear ahead of the rotational discontinuities.
  All physical quantities sharply change when crossing the slow shock as
  in the isotropic case.
  Although both parallel and perpendicular pressures increase across
  the shock, the parallel pressure is preferentially enhanced.
  It is notable that the temperature calculated from the perpendicular
  pressure slightly decreases, which implies that the required entropy
  increase across the shock is predominantly provided in the parallel
  direction.
  The anisotropy parameter,
  $\varepsilon=1-4\pi\left(p_{||}-p_{\perp}\right)/B^2$, is about 0.43 
  in this $\phi=30^\circ$ case.

\begin{figure}
   \includegraphics{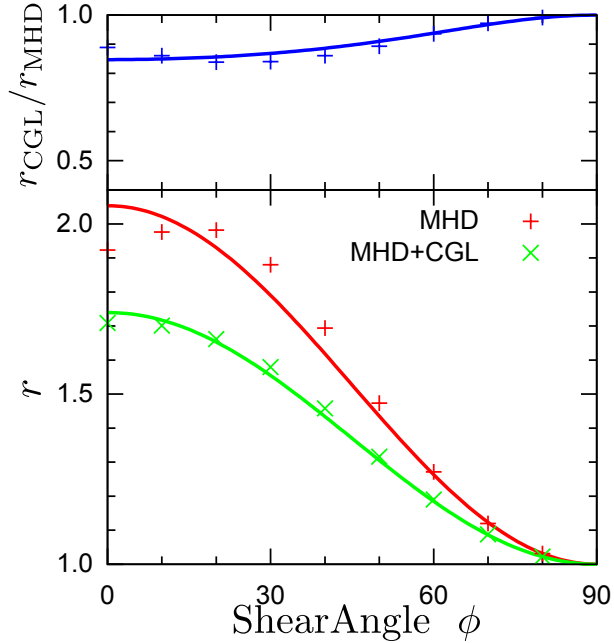}
   \caption{\label{fig:figure3}Shear angle dependence of the
   compression ratio of plasma density across the slow shocks.
   In the bottom panel, two types of points indicate 
   $r=\rho_{\rm down}/\rho_{\rm up}$ measured in 1-D
   simulations of isotropic and anisotropic MHD. 
   Solid lines are fitting functions with the form of $a\cos^2\phi+1$.
   The top panel shows the ratio of $r$ in the anisotropic MHD to that
   in the isotropic MHD.}
  \end{figure}

  The compression ratio of plasma density from the lobe to the inner
  plasma sheet becomes 1.58, which is less than that in the isotropic
  MHD case by about 16\%.
  The shear angle dependence of the compression ratio is summarized in
  Fig. \ref{fig:figure3}.
  The plus and cross symbols indicate the MHD and MHD+CGL
  cases, respectively.
  They are fitted by $a\cos^2\phi+1$ curves. 
  The top panel shows the ratio of compression ratio in anisotropic
  calculations to that in the isotropic ones, and the solid line
  indicates the ratio obtained by the $a\cos^2\phi+1$ curve fit in the
  bottom panel.
  A common feature in a wide range of shear angles is that the
  compression ratio decreases in an anisotropic MHD.
  In other words, the slow shock in the anisotropic MHD is always
  weaker than that in the isotropic MHD.
  The ratio weakly depends on the shear angle, but it has a value from
  0.8 to 1.0.

  The most striking modification from the isotropic case is that the
  slow shocks are formed ahead of the rotational waves.
  It is known that under the presence of the pressure anisotropy, the
  phase speed of an ${\rm Alfv\acute{e}n}$ wave is reduced to the value
  $V_A^* = \sqrt{\varepsilon} V_A$.
  The phase speed of slow modes, on the other hand, increases for an
  oblique propagation to the background magnetic field.
  Such a correction of each phase velocity modifies the structure of the
  reconnection layer.
  The antecedent of slow shocks can also be verified in the hodogram
  in Fig. \ref{fig:figure2}(b).
  The reduction of the magnetic field (1-2, 1'-2') occurs before the
  rotational waves (2-3, 2'-3').
  The same behavior can be seen in the case of $\phi=10^\circ$.
  In the two panels of $\phi=50^\circ$ and $\phi=70^\circ$ on the right,
  on the other hand, the rotation appears in the lobe side.

\begin{center}
   \begin{figure*}
   \includegraphics{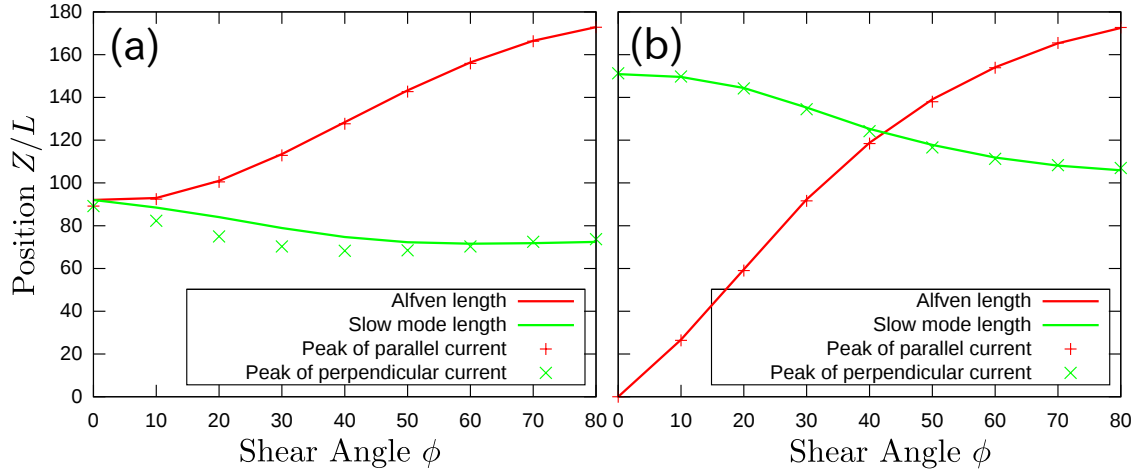}
   \caption{\label{fig:figure4}The relation between the shear angle
   $\phi$ and the wave-front of each mode. The positions at the peak
   parallel and perpendicular current density can be regarded as the
   ${\rm Alfv\acute{e}n}$ wave-front and the slow wave-fronts
   respectively. The solid lines are the theoretical propagation length
   obtained from simulation data. (a) shows MHD, and (b) MHD+CGL.}
   \end{figure*}
  \end{center}

  The relation between the shear angle $\phi$ and the positions of
  wave-fronts are summarized in Fig. \ref{fig:figure4}, for
  (a)MHD cases, and (b)MHD+CGL cases.
  Both the plus symbols and the red lines represent the front position
  of the ${\rm Alfv\acute{e}n}$ waves, while the cross symbol and green
  lines are the slow mode waves.
  The plus and cross symbols indicate respectively the positions at
  the peak parallel or perpendicular current density measured in our
  simulation.
  Since an ${\rm Alfv\acute{e}n}$ wave only rotates in the direction of
  the magnetic field, it can generate current only parallel to the field
  line.
  Hence, the location at the peak of the parallel current profile can
  be regarded as the ${\rm Alfv\acute{e}n}$ wave front, and the location
  where the perpendicular current becomes largest can be regarded as the
  slow-mode wave front.
  The solid lines represent the theoretical distance estimated by the product
  of the lapsed time and the phase velocity calculated from
  simulation data behind each wave front.
  The good agreement between the solid lines and dots confirms that the
  rotation and reduction of the magnetic field are indeed caused by
  ${\rm Alfv\acute{e}n}$ waves and slow-mode waves, respectively.
  In the anisotropic case, as the shear angle increases, the anisotropy
  parameter $\varepsilon$ arising downstream approaches the isotropic
  value, i.e., $\varepsilon \to 1$.
  Hence, as described above, the modified ${\rm Alfv\acute{e}n}$ velocity
  $V_A^*$ becomes larger as increasing the shear angle $\phi$ increases.
  The slow-mode velocity, on the other hand, decreases with $\phi$.
  As a result, rotational waves forego slow shocks for the shear angle
  greater than about $43^\circ$, while the rotational waves always form
  in front of the slow shocks in the isotropic MHD.

\begin{center}
   \begin{figure}
    \includegraphics{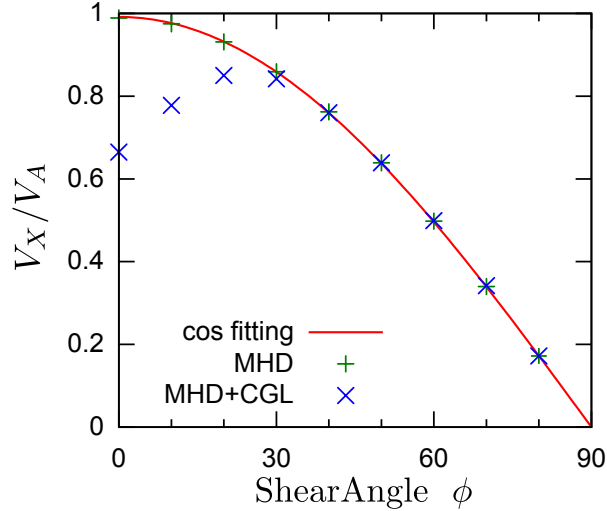}
    \caption{\label{fig:figure5}The dependence of outflow velocity upon
    the shear angle $\phi$. The plus and cross symbols indicate
    respectively $V_X$ measured at $Z=0$ in isotropic and anisotropic
    MHD, and the solid curve shows the ${\rm Alfv\acute{e}n}$ velocity
    by anti-parallel magnetic field component, i.e., $V_A\cos\theta$.}
   \end{figure}
  \end{center}

  Next, we shall discuss the relation between the in-/out-flow
  velocities and the guide field.
  The dependence of outflow velocities upon the shear angle $\phi$ is
  quite different between isotropic and anisotropic cases, particularly
  for the cases when slow shocks precede rotational waves.
  The outflow speed in the 1-D simulation is shown in
  Fig. \ref{fig:figure5}. 
  The plus and cross symbols indicate respectively $V_X$ measured at
  $Z=0$ in the MHD and MHD+CGL calculations, and the solid curve shows
  the 
  ${\rm Alfv\acute{e}n}$ velocity defined by an anti-parallel magnetic
  field component, i.e., $V_A\cos\theta$.
  The plus symbols of the isotropic MHD agree well with the
  ${\rm Alfv\acute{e}n}$ velocity measured by an anti-parallel magnetic
  field.
  For the anisotropic or MHD+CGL case, however, the outflow velocity
  becomes 
  smaller than the ${\rm Alfv\acute{e}n}$ speed for
  $\phi \ltsim 30^\circ$, at which slow shocks propagate faster than
  rotational waves.
  For a somewhat small $\phi$, the plasma flowing into a reconnection
  layer undergoes a two-step acceleration toward the $X$-direction,
  first by a slow shock, and then by a rotational wave.
  Since the released magnetic energy across a slow shock is very small
  in the anisotropic case, the resultant velocity across the shock
  becomes slower compared with the ${\rm Alfv\acute{e}n}$ velocity.
  For the case of $\phi \gtsim 40^\circ$, a rotational wave propagates
  faster than a slow-mode wave, and then the final $V_X$ is determined
  only by the rotational discontinuity, and a slow shock just reduces
  $V_Y$. 
  Thus, the sequence of wave propagation can control the nature of
  reconnection exhaust.

\begin{figure}
   \includegraphics{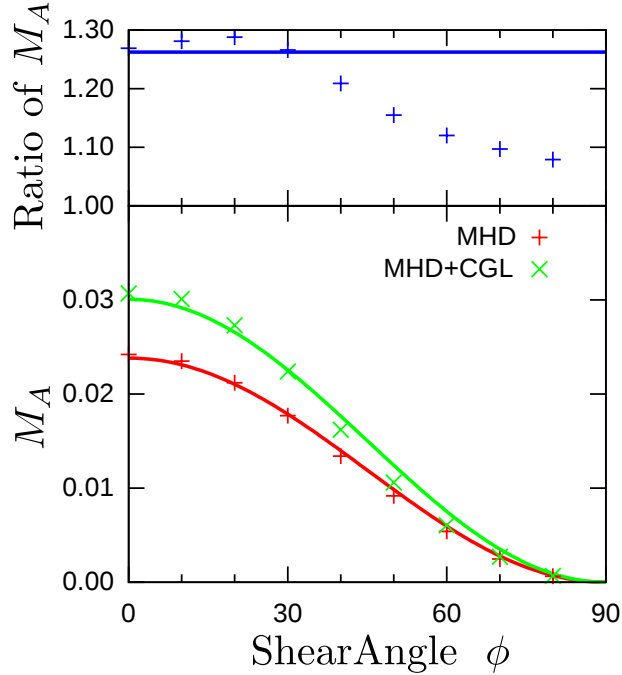}
   \caption{\label{fig:figure6}Shear angle dependence of inflow
   ${\rm Alfv\acute{e}n}$ Mach numbers, i.e., reconnection rates.
   In the bottom panel, two types of points indicate $V_Z/V_A$ measured at
   $Z/L=200$ in 1-D simulations of isotropic and anisotropic MHD.
   Solid lines are fitting functions proportional to $\cos^2\phi$.
   The top panel shows the ratio of reconnection rate in the anisotropic MHD
   to that in the isotropic MHD.} 
  \end{figure}

  We next discuss the reconnection rate, $M_A = V_{\rm in} / V_A$,
  defined by the inflow velocity and an upstream ${\rm Alfv\acute{e}n}$
  velocity measured at $Z/L=200$. 
  The relation between the reconnection rates and the shear angle
  surveyed by using 1-D simulations is summarized in
  Fig. \ref{fig:figure6}. 
  The notation of the symbols and solid lines are all same as
  Fig. \ref{fig:figure3}, but the fitting curves have the functional forms of
  $a\cos^2\phi$ now.
  For a wide range of shear angles, reconnection rate slightly rises in
  the MHD+CGL cases.
  The ratio of the reconnection rates has a value from 1.1 to 1.3.

  \subsection{\label{subsec:2d}2-D Simulation with Guide Field}

\begin{center}
   \begin{figure*}
    \includegraphics{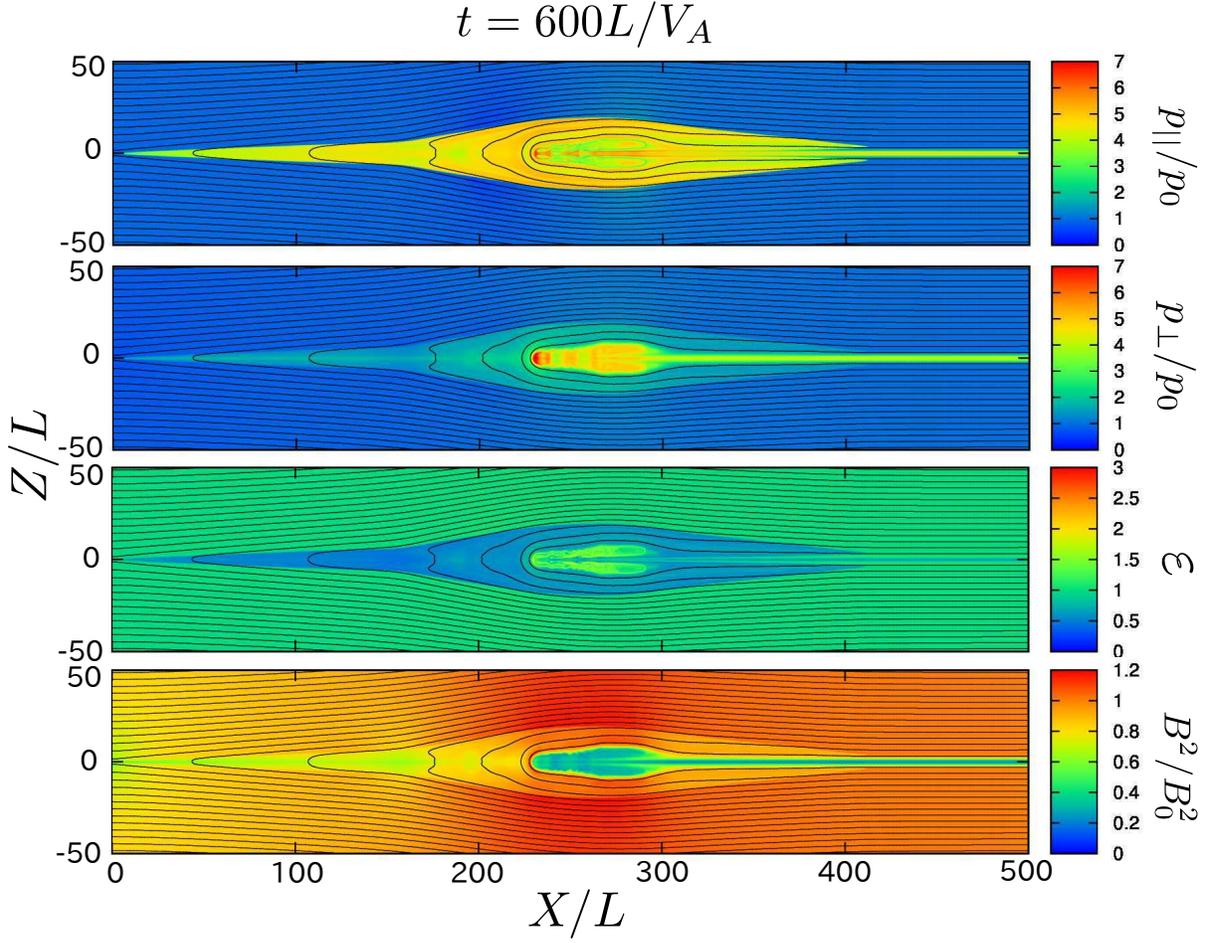}
    \caption{\label{fig:figure7}A snapshot of the MHD+CGL
    simulation with $\phi=30^{\circ}$ at time $t=600L/V_A$.
    Each panel shows the color contour of the parallel pressure, the
    perpendicular pressure, the anisotropy parameter, and the magnetic
    energy from top to bottom.
    All variables are normalized by the initial values in the lobe
    region. 
    The solid lines indicate the magnetic field lines.}
   \end{figure*}
  \end{center}

\begin{center}
   \begin{figure*}
    \includegraphics{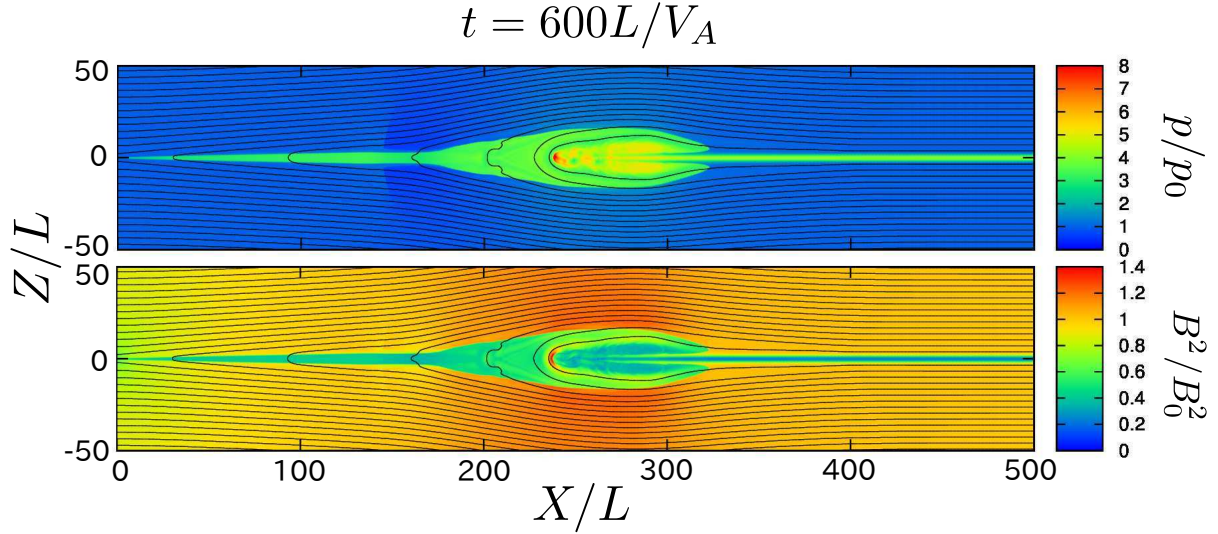}
    \caption{\label{fig:figure8}A snapshot of the isotropic MHD
    simulation with $\phi=30^{\circ}$ at time $t=600L/V_A$.
    The top and bottom panels show the color contours of the pressure
    and the magnetic energy, respectively, normalized by the initial
    values in the lobe region.
    The solid lines indicate the magnetic field lines.}
   \end{figure*}
  \end{center}

  With the results of the 1-D simulation fully in mind, we shall now
  turn to the results of the 2-D simulations.
  Fig. \ref{fig:figure7} shows a typical case for a developed
  reconnection layer 
  with the shear angle $\phi=30^{\circ}$ at time $t=600L/V_A$
  for an anisotropic MHD calculation, i.e., a MHD + CGL + Landau Closure
  calculation. 
  Each panel shows the color contours of $p_{||}$, $p_{\perp}$,
  $\varepsilon$, and $B^2$, normalized by the initial values in the lobe
  region respectively.
  Magnetic field lines are superposed by solid black lines.
  The results from an isotropic MHD calculation, the pressure and the
  magnetic energy, are shown in Fig. \ref{fig:figure8}.

  Once magnetic reconnection occurs, a high pressure plasmoid consisting
  of the dense plasma that initially supports the plasma sheet is
  squeezed out to reconnection exhaust.
  In the post-plasmoid plasma sheet after the passage of the plasmoid,
  the parallel pressure in the outflow region is enhanced more intensively
  than the perpendicular pressure across a pair of the weak slow shock
  boundaries, as we have already seen in 1-D simulations.
  The magnetic energy is reduced in the reconnection layer, but the
  reduction occurs less sharply than that in the isotropic case
  (see the bottommost panels in Fig. \ref{fig:figure7} and
  Fig. \ref{fig:figure8}).
  The downstream anisotropy parameter $\varepsilon \sim 0.45$ well
  agrees with the corresponding 1-D calculation.
  It is notable that the flaring angle sandwiched by two slow shocks
  becomes wider than the angle in an isotropic MHD simulation started
  from the same initial condition.
  For the shear angle $\phi=30^{\circ}$, these angles are $2.4^{\circ}$
  in the MHD+CGL and $1.4^{\circ}$ in the MHD.

\begin{figure}
   \includegraphics{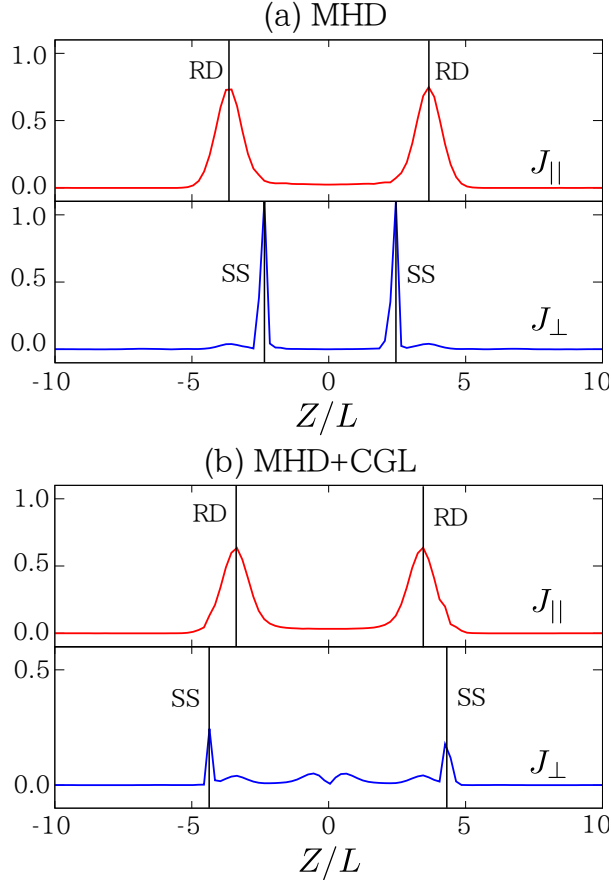}
   \caption{{\label{fig:figure9}}A cross-sectional view of current
   density at time $t=600L/V_A$ along $X=100L$.
   The top and bottom panels show the components parallel and
   perpendicular to the background magnetic field, respectively.
   (a) shows MHD, and (b) MHD + CGL.}
  \end{figure}

  To illustrate the spatial distribution of the reconnection layer, a
  cross-sectional view of the current density along the $Z$-axis at
  $X=100L$ is shown in Fig. \ref{fig:figure9}, (a) showing the isotropic
  MHD case, and (b) the anisotropic case. 
  The top and bottom panels indicate the current
  density parallel and perpendicular to the background magnetic field,
  normalized by the initial field in the lobe region divided by the half
  width of the initial current sheet, $B_0/L$.
  The vertical lines indicate the locations where the current densities
  reach their maximum. 
  Since $J_{||}$ and $J_{\perp}$ arise in rotational discontinuities and
  slow shocks respectively, as mentioned in the previous section, the
  vertical lines are marked by the scripts RD and SS.
  In fact, the rotational wave occupies a wider region here, and the
  slow shocks may be disguised.
  From the locations with the peak current density, however, it is clear
  that the slow shocks form in front of the rotational waves in the
  anisotropic case.
  Furthermore, the magnitude of the perpendicular current in the
  MHD+CGL is about a quarter of that in the isotropic MHD,
  which implies that the jump of the magnetic field across the slow
  shock is relatively small, and that the slow shock is weak from the
  point of view of the change of the magnetic field.

\begin{figure}
   \includegraphics{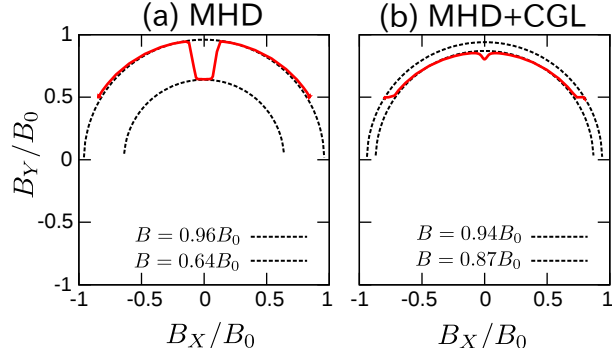}
   \caption{{\label{fig:figure10}}Hodograms of the magnetic
   field at time $t=600L/V_A$ along $X=100L$.
   The solid lines indicate simulation data and the dashed lines show
   equi-magnetic-energy circles.
   (a) is the isotropic MHD case and (b) the anisotropic MHD case (MHD +
   CGL).} 
  \end{figure}

  We can also analyze hodograms of the magnetic field along the $Z$-axis
  at $X=100L$ in Fig. \ref{fig:figure10}.
  The hodograms in (a)MHD and (b)MHD+CGL calculations are
  obtained from the same initial condition with $\phi=30^{\circ}$.
  The solid lines indicate simulation data, and the dashed lines
  equi-magnetic-energy circles.
  The outer and inner circles in both panels correspond to the magnetic
  energy upstream and downstream of slow shocks.

  In the isotropic MHD case, the rotational waves rotate the magnetic
  field 
  until $B_X$ becomes nearly equal to zero, and the slow shocks release
  the magnetic energy through the reduction of $B_Y$.
  In the MHD+CGL case, on the other hand, the slow shocks occur
  before the rotation of the magnetic field begins.
  Furthermore, the amount of magnetic energy released across the shocks
  in the MHD+CGL case (i.e., the width between the two
  equi-magnetic-energy circles) is much smaller than that in MHD case.
  These features are consistent with the steady state reconnection layer
  predicted by the 1-D Riemann problem (See Fig. \ref{fig:figure2}).

\begin{center}
   \begin{figure*}
    \includegraphics{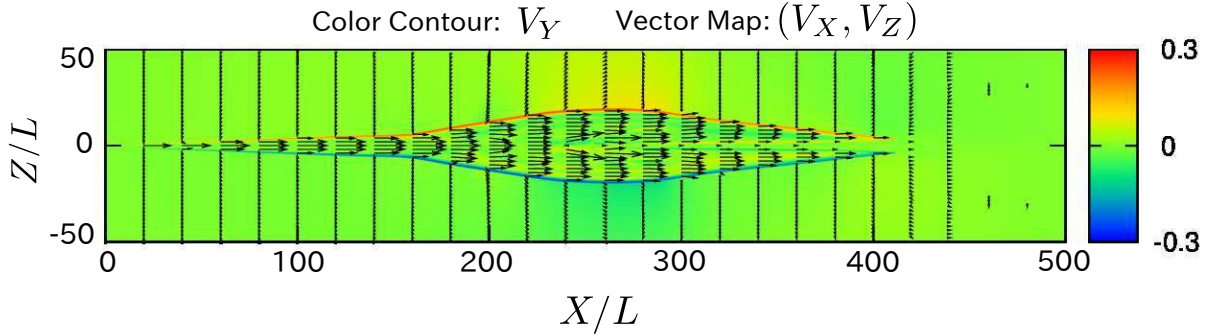}
    \caption{\label{fig:figure11}A snapshot of the MHD+CGL
    simulation with $\phi=30^{\circ}$ at time $t=600L/V_A$.
    The color contour shows the out-of-plane velocity, $V_Y$, normalized
    by the ${\rm Alfv\acute{e}n}$ velocity in the lobe region, $V_A$,
    and the vector map shows the in-plane components, $(V_X, V_Z)$.
    The length of an arrow of $20L$ corresponds to $1V_A$.}
   \end{figure*}
  \end{center}

  The velocity profile is also modified by the pressure anisotropy.
  A snapshot of the flow velocity at time $t=600L/V_A$ is shown in
  Fig. \ref{fig:figure11}.
  The color contour indicates the out-of-plane velocity, $V_Y$, and the
  vector map signifies the in-plane velocity vector, $(V_X, V_Z)$,
  normalized by $V_A$.
  The length of an arrow of $20L$ corresponds to $1V_A$.
  The boundary layers where the finite $V_Y$ arise are almost identical to
  the positions of rotational discontinuities.
  The plasma advected from the lobe region is accelerated to the 
  $X$-direction across the boundary layer until an order of the
  ${\rm Alfv\acute{e}n}$ velocity.
  The outflow speed measured at $\left(X,Z\right)=\left(100L,0\right)$
  reaches $0.76V_A$.
  In our isotropic MHD simulation (not shown here), the outflow speed
  measured at $(X,Z)=(100L,0)$ reaches $0.78V_A$, which is slightly
  less than the value $V_A\cos\phi = 0.86V_A$, 
  The outflow speed in the MHD+CGL case is smaller than an ordinary
  ${\rm Alfv\acute{e}n}$ velocity, $V_A\cos\phi$, and than that in the
  isotropic MHD case, but larger than a modified ${\rm Alfv\acute{e}n}$
  velocity, $V_A^* \cos\phi= \sqrt{\varepsilon}V_A \cos\phi= 0.58V_A$
  calculated from the downstream anisotropy parameter.
  In 1-D simulations, it is assumed that all variables are uniform in
  the $X$-direction, but here in 2-D cases, non-uniformity along $X$-axis
  somewhat reduces the outflow velocities both in the isotropic and
  the anisotropic cases.

  By measuring the inflow velocity at the point $(X,Z)=(0,50L)$,
  we found that the reconnection rates are $M_A = 0.028$ in the
  MHD case, and $M_A = 0.035$ in the MHD+CGL case.
  The ratio between these reconnection rates is $1.25$, and this is
  consistent with 1-D simulation results, discussed in previous section
  (see Fig. \ref{fig:figure6}). 

  \section{\label{sec:discussion}Discussion and Conclusion}

  We summarize the main results obtained from the 1-D and 2-D
  simulations as follows:
  \begin{enumerate}
   \item
	Once magnetic reconnection takes place, a pair of slow shocks
	does form even in an anisotropic MHD.
   \item
	The parallel pressure increases mainly across the slow shock,
	and the large pressure anisotropy is induced.
   \item
	The generated slow shocks are much weaker than those formed in an
	isotropic MHD, from the aspects of the compression ratio and the
	amount of released magnetic energy.
   \item
	In spite of the weakness of the slow shocks, the resultant
	reconnection rate is $10-30\%$ larger than that in an isotropic
	MHD.
  \end{enumerate}

  Result 1 is verified by comparing the downstream variables with
  theoretical values predicted by the Rankine-Hugoniot relations in
  anisotropic plasmas,\cite{Karimabadi1995,Chao1970,Hudson1971} in which
  a downstream anisotropy parameter $\varepsilon_2$ is treated as a given
  parameter obtained from the simulation data.
  From the point of view of the fluid approximation, as Petschek pointed
  out, magnetically stored energy is actually converted to thermal and
  kinetic energy in plasmas through the formation of a pair of slow
  shocks even under the pressure anisotropy. 
  The modification from the reconnection layers in isotropic fluids is
  mentioned in Results 2-4.

  Result 2 is a straightforward outcome of the double adiabatic
  theory, as mentioned in Section \ref{sec:model}.
  This result may correspond to the microscopic view of the plasma sheet
  boundary layer, such that the PSBL ion beam accelerated along the
  background magnetic field distorts the distribution function in kinetic
  regime.\cite{Higashimori2012}
  The second order momentum of the distribution function suggests
  $p_{||} > p_{\perp}$.

  The downstream anisotropy parameter, $\varepsilon_2$, strongly depends
  on both the shear angle of the field lines, $\phi$, and the upstream
  plasma beta, but when $\phi$ approaches $0^\circ$, $\varepsilon_2$
  becomes nearly $0$, which corresponds to the marginal firehose
  criterion.
  According to previous kinetic simulations,
  \cite{Liu2011a,Liu2011b} however, the downstream anisotropy parameter 
  is somewhat larger than $0$.
  These authors argue that the ${\rm Alfv\acute{e}nic}$
  counter-streaming ions, here corresponding to the parallel pressure
  enhancement, drive the drop of $\varepsilon_2$, and that downstream
  turbulent waves with the ion inertial scale scatter particles and
  raise $\varepsilon_2$.
  Since there is no counterpart of inertial scale waves in our
  simulations, it is consistent that $\varepsilon_2$ falls to a
  value smaller than that in the kinetic simulations.
  Comparisons between the 1-D and 2-D calculations show that
  $\varepsilon_2$ obtained in the 2-D case is slightly larger than the
  corresponding 1-D value, at least for $\phi$ larger than $20^\circ$.
  The multi-dimensionality, or non-uniformity in the $X$-direction, may
  also be essential for the development of the anisotropy in
  reconnection.

  Results 3 and 4 are the most significant results in this
  work.
  Despite the smallness of the amount of magnetic energy released only
  across the slow shocks, the reconnection rates are maintained or
  slightly increase in the MHD+CGL calculations.
  Due to a relatively small enhancement of perpendicular
  pressure, pressure equilibrium related to $p_{\perp} + B^2/8\pi$
  cannot be maintained, and the current sheet is squashed.
  This unbalance tends to enlarge the inflow velocities and the
  reconnection rates.
  Furthermore, the acceleration of reconnection exhaust is realized
  not only by the weak slow shocks, but also by the rotational
  discontinuities, which is striking especially for relatively small
  guide field reconnection.
  Note that the similar results have been obtained in the isotropic MHD
  \cite{Tsai2006}.
  
  Finally, our results may be able to explain the rareness of the
  {\it in-situ} observations of the Petschek-type reconnection
  accompanied by slow shocks.
  The slow shocks formed in a collisionless magnetic reconnection
  might be too weak to be observed in their own right.
  Moreover, not only in the isotropic MHD\cite{Tsai2006}, but also in
  the anisotropic MHD, the results imply that the slow shocks do not
  necessarily play an important role in the energy conversion in the
  collisionless magnetic reconnection system.
  
  \begin{acknowledgments}
   We thank Takanobu Amano, Yoshifumi Saito, and Katsuaki Higashimori
   for useful discussions.
   This work was supported by the editing assistance from the GCOE
   program.
  \end{acknowledgments}

\end{document}